\documentclass[12pt]{article}
\usepackage{amsfonts}
\begin{document}
{\renewcommand{\thefootnote}{\fnsymbol{footnote}}
\hfill  PITHA -- 99/24\\
\medskip
\hfill hep--th/9907043\\
\medskip
\begin{center}
{\Large{The Area Operator in the Spherically Symmetric
Sector\vspace{0.2cm}\\ of Loop Quantum Gravity}}\\ \vspace{1cm}
{\large M.~Bojowald\footnote{e-mail address: {\tt
bojowald@physik.rwth-aachen.de}} and H.A.~Kastrup\footnote{e-mail
address: {\tt kastrup@physik.rwth-aachen.de}}} \vspace{0.3cm}\\
Institute for Theoretical Physics\\ RWTH Aachen, D-52056 Aachen,
Germany\\ \vspace{1cm}
\end{center}
\setcounter{footnote}{0}
\newtheorem{theo}{Theorem}
\newtheorem{lemma}[theo]{Lemma}
\newtheorem{defi}{Definition}
\newcommand{\be}{\begin{equation}}
\newcommand{\ee}{\end{equation}}
\newcommand{\proofend}{\raisebox{1.3mm}{\fbox{\begin{minipage}[b][0cm][b]{0cm}
\end{minipage}}}}
\newenvironment{proof}{\noindent{\it Proof:} }{\mbox{}\hfill \proofend\\\mbox{}}
\newenvironment{ex}{\noindent{\it Example:} }{\medskip}
\newenvironment{rem}{\noindent{\it Remark:} }{\medskip}

\newcommand{\Ab}{\overline{{\cal A}}}
\newcommand{\Gb}{\overline{{\cal G}}}
\newcommand{\AGb}{\overline{{\cal A/G}}}
\newcommand{\Ub}{\overline{{\cal U}}}
\newcommand{\AbR}{\overline{{\cal A}}_B}
\newcommand{\GbR}{\overline{{\cal G}}_B}
\newcommand{\AGbR}{\overline{{\cal A/G}}_B}
\newcommand{\UbR}{\overline{{\cal U}}_B}
\newcommand{\AbS}{\overline{{\cal A}}_{\Sigma}}
\newcommand{\GbS}{\overline{{\cal G}}_{\Sigma}}
\newcommand{\AGbS}{\overline{{\cal A/G}}_{\Sigma}}
\newcommand{\UbS}{\overline{{\cal U}}_{\Sigma}}
\newcommand{\TAb}{\overline{{\cal T}}}
\newcommand{\PAb}{\overline{{\cal P}}}
\newcommand{\Ag}{{\cal A}_{\gamma}}
\newcommand{\Gg}{{\cal G}_{\gamma}}
\newcommand{\Ug}{{\cal U}_{\gamma}}
\newcommand{\Lie}{{\cal L}}
\newcommand{\pgg}{p_{\gamma\gamma^{\prime}}}
\newcommand{\Phip}{\Phi^{\prime}}
\newcommand{\lp}{\lambda^{\prime}}
\newcommand{\rstar}{\overline{r}_{[\lambda]}^{\star}}
\newcommand{\rnstar}{\overline{r}_{[\lambda_n]}^{\star}}
\newcommand{\sigl}{\sigma_{[\lambda]}}
\newcommand{\sigln}{\sigma_{[\lambda_n]}}
\newcommand{\kt}{\vartheta}
\newcommand{\kp}{\varphi}
\newcommand{\Eg}{E(\gamma)}
\newcommand{\Vg}{V(\gamma)}

\newcommand{\md}{\mathchoice%
  {\mbox{\rm d}}%
  {\mbox{\rm d}}%
  {\mbox{\scriptsize\rm d}}%
  {\mbox{\tiny\rm d}}}
\newcommand{\td}[2][]{\mathchoice%
  {\frac{\mbox{\small\rm d}#1}{\mbox{\small\rm d}#2}}%
  {\frac{\mbox{\scriptsize\rm d}#1}{\mbox{\scriptsize\rm d}#2}}%
  {\frac{\mbox{\tiny\rm d}#1}{\mbox{\tiny\rm d}#2}}%
  {\frac{\mbox{\tiny\rm d}#1}{\mbox{\tiny\rm d}#2}}}
\newcommand{\pd}[2][]{\frac{\partial #1}{\partial #2}}
\newcommand{\fd}[2][]{\frac{\delta #1}{\delta #2}}
\newcommand{\Aut}{\mbox{\rm Aut}\,}
\newcommand{\Ad}{\mbox{\rm Ad}\,}
\newcommand{\ad}{\mbox{\rm ad}\,}
\newcommand{\Hom}{\mbox{\rm Hom}\,}
\newcommand{\Ima}{\mbox{\rm Im}\,}
\newcommand{\id}{\mbox{\rm id}\,}
\newcommand{\diag}{\mbox{\rm diag}\,}
\newcommand{\Kern}{\mbox{\rm ker}\,}
\newcommand{\tr}{\mbox{\rm tr}\,}
\newcommand{\sgn}{\mbox{\rm sgn}\,}

\newcommand{\DirB}[3][\mkern-5mu]{\ensuremath{%
 \left\langle#2\left|#1\right|#3\right\rangle}}
\newcommand{\bra}[1]{\ensuremath{\left\langle#1\right|}}
\newcommand{\ket}[1]{\ensuremath{\left|#1\right\rangle}}

\newcommand*{\R}{{\mathbb R}}
\newcommand*{\N}{{\mathbb N}}
\newcommand*{\Z}{{\mathbb Z}}
\newcommand*{\Q}{{\mathbb Q}}
\newcommand*{\C}{{\mathbb C}}

\begin{abstract}
Utilizing the previously established general formalism for quantum
symmetry reduction in the framework of loop quantum gravity the
spectrum of the  area operator acting on spherically symmetric
states in 4 dimensional pure gravity is investigated.

The analysis requires a careful treatment of partial
 gauge fixing  in the classical symmetry reduction and
of the reinforcement of $SU(2)$-gauge invariance for the
quantization of the area operator.
 The  eigenvalues of that operator applied to the spherically
 symmetric spin network states have the form
 $$A_n\propto\sqrt{n(n+2)},~n=0,1,2,\ldots~,
 $$ giving $A_n \propto n$ for large $n$.

The result clarifies (and reconciles!) the relationship between
the more complicated spectrum of the general (non-symmetric) area
operator in loop quantum gravity and the old Bekenstein proposal
that $A_n \propto n$.
\end{abstract}

\section{Introduction}

In a preceding paper \cite{bo1} we proposed a framework for
implementing symmetry reductions for gravitational systems
quantized within the loop quantum gravity approach (see the
reviews \cite{ro1}). In the present note we apply that framework
to the spectrum of the operator associated with 2-dimensional
areas. The problem is of considerable physical interest because
the 2-dimensional horizon of a Schwarzschild black hole
constitutes such a system the area of which is, up to a factor, a
measure for the entropy of the black hole.

The quantum area spectrum of the horizon of (Schwarzschild) black
holes in 4-dimensional space-time has a longer history: Already in
1974 Bekenstein, using Bohr-Sommerfeld type arguments \cite{be1},
suggested that the area $A=4\pi R_S^2,~ R_S=2GM^2/c^2$ of a
(spherically symmetric) Schwarzschild black hole of mass $M$ has
an angular momentum like quantum area spectrum, $A(n)\propto n,~ n
\in \N \equiv \{n=1,2,\ldots \}$, yielding an energy spectrum
$E_n\propto \sqrt{n}$. In the meantime such a spectrum has been
argued for by many authors (for details and the corresponding
literature see Refs.\ \cite{be2,ka2,ka5}).

 A very
recent group theoretical quantization based on the classical
canonical structure of the Schwarzschild system in $D~(\geq 4)$
space-time dimensions  and the group $SO^{\uparrow}(1,2)$ yields
the spectrum \cite{bo2} \be A_{D-2}(k;n)\propto (k+n),~n \in \N_0
\equiv \{n=0,1,2,\ldots\}~, \ee where $k$ characterizes the
irreducible unitary representation of $SO^{\uparrow}(1,2)$ or its
covering groups: For $SO^{\uparrow}(1,2)$ itself we have $k \in
\N$, for its two-fold coverings $SU(1,1) \cong SL(2,\R)$ $k \in
(1/2)\N$ and for the universal covering group $k$ may be any real
number $>0$.

On the other hand the spectrum of the general (non-symmetric) area
operator in loop quantum gravity is more complicated
\cite{ro2,ro3,asle}: Possible eigenvalues of the area operator in
this theory  are \be A\propto\sum_p\sqrt{j_p(j_p+1)}~, \ee where
$p$ labels points at which the surface is intersected by a spin
network and $j_p\in\frac{1}{2}\N_0$ is the spin of the edge
intersecting the surface in $p$. Here we have ignored the singular
case that the surface is intersected in a vertex of the spin
network.

There is an important difference between the spectra (1) and (2):
Whereas for the former the distance between successive eigenvalues
remains the same for any $n$ that  distance   becomes smaller and
smaller with increasing area for the spectrum (2) \cite{asle}.
This result has led to expressions of doubts
\cite{sm1,ro4,ro1,askr} as to the physical validity of the
spectrum (1) and its possible implications for the structure of
the semi-classical Hawking radiation \cite{bemu}.

Using the framework of Ref.\ \cite{bo1} we shall show how the
spectra (1) and (2) are related and how the two approaches are to
be reconciled:

An observation to start with is that the spectrum (2)  contains
the other spectrum (1) as a subset: If there are $n$ edges
intersecting the surface, all labeled with the same spin, e.g.\
$j=\frac{1}{2}$, then the area eigenvalue will be $A_n\propto n$.
In Ref.\ \cite{ABCK:LoopEntro} the horizon has been treated in
loop quantum gravity as a boundary with appropriate boundary
conditions which fix a direction in the internal $SU(2)$-space
thereby breaking $SU(2)$ to $U(1)$.  This effects a replacement of
the Casimir operator $J^2$ with eigenvalues $j(j+1)$ labeling
irreducible representations of $SU(2)$ by the parameter $n$
labeling irreducible representations of $U(1)$. According to Ref.\
\cite{kr1} this leads to a spectrum $A\propto \sum_p
j_p\in\frac{1}{2}\N_0$ which is again of a form of the old area
spectrum (1). Furthermore, it has been argued in Ref.\ \cite{kr1}
that the  loop quantum gravity area operator of Refs.\
\cite{ro2,asle} measures the area of  surfaces including {\em all}
fluctuations, with the restricting boundary condition, however,
that transversal fluctuations are suppressed at the horizon: The
direction of the normal to the surface cannot fluctuate.

Such a boundary condition  is analogous to the situation in a
symmetry reduced model: By imposing, e.g.\ spherical symmetry one
allows only for symmetric solutions and therefore only for
fluctuations respecting that symmetry.

In the present note we, accordingly,  analyze properties of the
area operator in a spherically symmetric sector of loop quantum
gravity along the lines suggested in \cite{bo1}. Analogously to
corresponding boundary conditions  at the horizon, the classical
symmetry reduction involves a reduction of the gauge group from
$SU(2)$ to $U(1)$. This is means a partial gauge fixing, which
will be undone in the quantum theory.

In the next section we shall summarize the essential steps
characterizing the classical spherical symmetry reduction and the
corresponding setup for the quantum theory. Section 3 contains an
analysis of the spectrum of the area operator respecting spherical
symmetry. It has the form \be A\propto\sqrt{j(j+1)}~~ \mbox{ (no
sum over punctures)}~. \ee Properties of that spectrum are
discussed in the final section.

\section{Symmetry Reduction \\ and Partial Gauge Fixing}

The quantum symmetry reduction of Ref.\ \cite{bo1} uses the
classification of symmetric principal fibre bundles and invariant
connections thereon \cite{KobNom,Harnad,Brodbeck}. In this
framework a symmetry group $S$ acts on a principal fibre bundle
$P(\Sigma,G,\pi)$ with compact structure group $G$ over the base
manifold $\Sigma$ which is a spacelike hypersurface of the
space-time used to carry out the canonical formalism. The symmetry
group is a finite-dimensional subgroup of the group of bundle
automorphisms, i.e.\ it commutes with the right action of $G$ on
$P$. Of special importance is the isotropy subgroup which here is
assumed  to be the same for all points in $P$ (if not, the
  base manifold can be decomposed into components all having the same
  isotropy subgroup.  This amounts to cutting out symmetry centers and
  axes).

  The isotropy subgroup $F<S$ of points in $P$ acts on each fibre and
therefore determines a homomorphism $\lambda_p\colon F\to G$ by
$f(p)=:p\cdot\lambda_p(f)$ for all $f\in F$ and $p\in P$.
Homomorphisms $\lambda_p$ and $\lambda_{p\cdot g}$ related to
different points in the same fibre differ only by conjugation:
$\lambda_{p\cdot
  g}=\Ad_{g^{-1}}\circ\lambda_p$ for $g\in G$.   By the action
  of $S$ the base manifold $\Sigma$
becomes an orbit bundle $\Sigma\cong B\times S/F$ with base
manifold $B\cong \Sigma/S$ and orbits $S/F$. By choosing an
analytic section in this orbit bundle the base $B$ can be embedded
analytically in $\Sigma$. The bundle $P$ can be restricted to a
principal fibre bundle $P|_B$ over $B$, which can be further
reduced by defining $Q_{\lambda}:=\{p\in
P|_B:\lambda_p=\lambda\}$.  This reduction uses a fixed
homomorphism $\lambda\colon F\to G$, and the reduced bundles are
principal fibre bundles over $B$ with structure group
$Z_{\lambda}:=Z_G(\lambda(F))$, the centralizer in $G$ of
$\lambda(F)$. All symmetric principal fibre bundles $P$ are
classified by a conjugacy class $[\lambda]$ of homomorphisms and a
reduced bundle $Q$. As noted above, the homomorphisms $\lambda_p$
get conjugated by changing the point $p$ in the fibre. Therefore,
all homomorphisms in the conjugacy class $[\lambda]$ are
equivalent for classifying the symmetric bundle $P$; selecting one
of them amounts to a partial gauge fixing breaking the structure
group $G$ down to $Z_{\lambda}$.

Analogously an invariant connection $\omega$ on a symmetric fibre
bundle, classified by $\lambda$ and $Q_{\lambda}$, leads to a
$Z_{\lambda}$-connection on $Q_{\lambda}$ by restriction. To see
that fix a point $p\in P$ and a vector $v$ in $T_p P$. Then the
pull back of $\omega$ by $f\in F$ applied to $v$ is by definition
$f^{\star}\omega_p(v)=\omega_{f(p)}(\md f(v))$. If we now use the
fact that $f$ acts as gauge transformation in the fibres and
observe the definition of $\lambda_p$ and the adjoint
transformation of $\omega$, we obtain  $\omega_{f(p)}(\md
f(v))=\Ad_{\lambda_p(f)^{-1}}\omega_p(v)$ ($\omega_p$ annihilates
by definition the horizontal part of $v$, which only is changed by
$\md f$). By assumption the connection $\omega$ is $S$-invariant
implying
$f^{\star}\omega_p(v)=\Ad_{\lambda_p(f)^{-1}}\omega_p(v)=\omega_p(v)$.
This shows that $\omega_p(v)\in Z_G(\lambda_p(F))$, and $\omega$
can be restricted to a connection on the bundle $Q_{\lambda}$ with
structure group $Z_{\lambda}$.

 Besides the reduced connection on
$Q_{\lambda}$ there are several scalar fields, jointly denoted as
``Higgs'' field in the following, which stem from the components
of $\omega$ tangential to the $S$-orbits. The reduced connection
together with the Higgs field suffice to classify the invariant
connection completely: $\omega$ can be reconstructed out of this
data.  This observation was exploited in Ref.\ \cite{bo1} by using
the reconstruction map to pull back a cylindrical function on the
space of connections over $\Sigma$, i.e.\ a function in the
auxiliary Hilbert space of the unreduced gauge theory, to a
function on the space of connections and Higgs fields over $B$.
However, the reduction of the structure group had to be undone to
carry out the quantization procedure in the general framework. In
the following we will describe this in more detail for the example
of spherical symmetry.

We saw that the classification of symmetric bundles and invariant
connections makes use of a partial gauge fixing by choosing
$\lambda\in[\lambda]$. However, the full $G$-transformations are
implemented by acting on the classifying structure:
$Z_{\lambda}$-gauge transformations are gauge transformations on
the reduced bundle $Q_{\lambda}$; they constitute the reduced
gauge group. All elements of $G$ which are not contained in
$Z_{\lambda}$ change $\lambda$ by conjugation.  They change the
reduced bundles in an equivalence class, all of which yield the
same symmetric bundle after reconstruction.

From now on we specialize to $S=SU(2)=G$, $F=U(1)$ in order to
describe spherically symmetric solutions of general relativity in
the real Ashtekar formulation. In this case selecting one
$\lambda\in[\lambda]$ amounts to fixing an internal axis (a point
in $S^2$) and $Z_{\lambda}$-gauge transformations are rotations
around this axis, whereas a conjugation of $\lambda$ rotates the
axis. The group $G=SU(2)\cong S^3$ acting on $\lambda$ by
conjugation has the isotropy subgroup $Z_{\lambda}\cong U(1)$.
Factoring out this subgroup leads to the Hopf map $S^3\to S^2$.
The spacelike section $\Sigma$ of space-time is by the action of
$S$ an orbit bundle with $S^2$-fibres over the one-dimensional
base manifold $B$, which can be embedded in $\Sigma$ as a radial
manifold by choosing a section in the orbit bundle.

We can represent $F=\exp\langle\tau_3\rangle$ ($\langle \cdot
\rangle $ denotes the linear span) as subgroup of $S$, where
$\tau_j=-\frac{i}{2}\sigma_j$ are $SU(2)$-generators with the
Pauli matrices $\sigma_j$. Then all conjugacy classes of
homomorphisms $\lambda\colon F\to G$ are given by their
representatives
$\lambda_k\colon\exp(t\tau_3)\mapsto\exp(kt\tau_3), k\in\N_0$. For
$k\not=1$ they lead to degenerate sectors of vanishing volume (see
Ref.\ \cite{bo1} for details) and we will be mainly concerned with
the $(k=1)$-sector in the following. This is the only sector
exhibiting a non-vanishing Higgs field; the symmetry reduction
leads to the well known connections which are symmetric up to
gauge transformations \cite{CorderoTeit}.

The only $SU(2)$-gauge transformations fixing $\lambda_1$ are
generated by $\tau_3$, all other gauge transformations change
$\lambda$ in its conjugacy class. In this $\tau_3$-gauge we have
$\md\lambda_1(\tau_3)=\tau_3$, whereas in an arbitrary
$\lambda$-gauge, $\lambda\in[\lambda_1]$, we have
$\md\lambda(\tau_3)=g^{-1}\tau_3g,~g\in SU(2)$. In order to
characterize this general gauge by coordinates we parameterize
$SU(2)$ by Euler angles: $g=g_3(\psi)g_1(\kt)g_3(\kp)$ where
$g_i(t):=\exp t\tau_i$. This yields
\begin{eqnarray*}\md\lambda(\tau_3)&=&\sin\kt\sin\kp\,
\tau_1+\sin\kt\cos\kp\,\tau_2+
\cos\kt\,\tau_3=:n^i\tau_i~ \\ & &\mbox{with}~~n^i\,n_i\equiv
\vec{n}^2 =1,~n_i =\delta_{ij}n^j~.\end{eqnarray*} Fixed $\vec{n}$
corresponding to a fixed $\lambda\in[\lambda_1]$ is analogous to a
fixed direction in $SU(2)$ introduced in \cite{kr1}. However, as
shown above, a chosen $\vec{n}$ represents pure gauge and physical
states and observables should, of course, be independent of the
choice.

The classical phase space of the symmetry reduced theory consists
of fields $(A_I,E^I), 1\leq I\leq3$, where $A_1$ is the component
of the reduced connection over $B$, $E^1$ is its conjugate
momentum, and $A_2,A_3$ are Higgs field components with conjugate
momenta $E^2,E^3$. We will be concerned only with $A_1,E^1$
because the 2-dimensional area $A$ is classically given by \be A=
4\pi |E^1|~. \ee In the $\lambda$-gauge we have the
$U(1)$-connection form $$A_1\,n^i\tau_i\,\md x~,$$ where $x$ is a
(local) coordinate on $B$, and the ${\cal L}U(1)$-valued field
$E^1\,n^i\tau_i$. Their symplectic structure is given by \be
\{A_1(x),E^1(y)\}=\frac{\kappa\iota}{4\pi}\delta(x,y)\ee with
$\kappa=8\pi G$ and the Immirzi parameter $\iota$.

{\em Without} partial gauge fixing the fields would be
$SU(2)$-valued and given by $$A^i\tau_i\,\md x~~ \mbox{and}~~
E_i\tau^i$$  with
\begin{eqnarray*} A^1&=&A\sin\kt_A\sin\kp_A,~~
A^2=A\sin\kt_A\cos\kp_A,~~ A^3=A\cos\kt_A,\\ E_1&=&
E\sin\kt_E\sin\kp_E,~~ E_2=E\sin\kt_E\cos\kp_E,~~ E_3=E\cos\kt_E,
\end{eqnarray*}
in spherical coordinates and with  symplectic structure  \be
\{A^i(x),E_j(y)\}=\frac{\kappa\iota}{4\pi}\delta^i_j\delta(x,y)~.
\ee (The indices $I=1$ in Eq.\ (5) denote space indices whereas
the indices $i,j$ in Eq.\ (6) are $SU(2)$ indices which are
lowered or raised in terms of the Killing metric $(\delta_{ij})$.)
By using the spherical coordinates we can symplectically embed the
phase space $(A_1,E^1)$ in $\lambda$-gauge as a ray in the phase
space of $SU(2)$-valued fields: $A_1\mapsto A_1\,n^i=A^i$,
$E^1\mapsto E^1\,n_i=E_i$ with $\kt_A=\kt_E$ and $\kp_A=\kp_E$
fixed such that the direction of $n^i$ is given by the angles
$\kt_A,\kp_A$.

{\em If} starting quantization from the phase space $(A_1,E^1)$ we
would arrive at $U(1)$-spin networks.  However, this renders the
partial gauge fixing manifest and even worse, a Higgs field cannot
be included easily in Higgs vertices using the framework of Ref.\
\cite{FermionHiggs} because it transforms in the adjoint
representation of $SU(2)$, not of $U(1)$ which is, of course, the
trivial representation. As described in Ref.\ \cite{bo1} we can
undo the partial gauge fixing in the quantum theory by extending
the spin networks by $SU(2)$-gauge invariance to spin network
functions on the space of $SU(2)$-connections and appropriate
Higgs fields over $B$.  The spin network functions then depend not
only on $A_1$ (and Higgs field components) but on all
$SU(2)$-components $A^i$ introduced above. Accordingly, the
$SU(2)$-components $E_i$ get quantized to functional derivatives
$$\hat{E}_i(x)=\frac{\hbar\kappa\iota}{4\pi i}\fd{A^i}(x)~.$$ This
will be crucial for the  quantization of the area operator.

Before addressing that problem we mention that the extension  to
$SU(2)$-invariant spin networks can be understood as integrating
the partial gauge fixings $\vec{n}$ over $S^2$ for any edge in the
graph underlying the spin network state.  In order to make this
precise we use coherent states on $SU(2)$ introduced in Ref.\
\cite{Coherent}: These are defined for a fixed irreducible
representation of $SU(2)$ with weight $j$ and a state $\ket{j,m}$
therein by
\begin{equation}\label{coherent}
  \ket{m,\vec{n}}_j:=\pi^{(j)}(g_3(\kp)g_1(\kt))\ket{j,m}~,
   ~~\mbox{for all}~~ \vec{n} \in S^2.
\end{equation}
Here $\pi^{(j)}$ is the irreducible $SU(2)$-representation of
weight $j$, and $\kt,\kp$ are coordinates of $\vec{n}$ in $S^2$.
These coherent states are (over-)complete for any fixed $j,m$,
namely
\begin{equation}\label{complete}
  \frac{2j+1}{4\pi}\int_{S^2}\md^2n\ket{m,\vec{n}}_j
   \bra{m,\vec{n}}_j=\pi^{(j)}(1)
\end{equation}
is the identity operator in the $j$-representation. We can now
project each edge holonomy in an $SU(2)$-spin network to a
$U(1)$-holonomy by inserting the projector
$\ket{m,\vec{n}}_j\bra{m,\vec{n}}_j$ in each edge with spin $j$,
where $m\, ;-j\leq m\leq j$ is arbitrary but fixed (for each $j$).
At this point there arises an arbitrariness because any
$SU(2)$-representation splits into several representations
(labeled by $m$) of a $U(1)$-subgroup. This yields the holonomies
of a $U(1)$-spin network in the $\lambda$-gauge where $\lambda$
can be chosen arbitrarily for each edge (in the classification one
uses only $\lambda$ which are constant along $B$ for simplicity.
Such a choice is always possible by choosing an appropriate
section in $P|_B$. However, $\lambda$ is defined by the action of
$F$ in each point of $P$ and can, of course, vary in different
points). Note, however, that we have no such projection procedure
for Higgs vertices; a projection of a full spin network and,
therefore, a partial gauge fixing in the quantum theory can
completely be done only in the degenerate sectors which have no
Higgs vertices \cite{bo1}.

  Arrived at a $U(1)$-spin network
(and disregarding Higgs vertices), we can multiply the
corresponding states for each edge with $(4\pi)^{-1}(2j+1)$ and
integrate $\vec{n}$ over $S^2$. Using the completeness relation
(\ref{complete}) we see that we get back the original $SU(2)$-spin
network.

\section{Area Operator}
Before quantizing the (spherically symmetric) area operator
without gauge fixing let us first rephrase the results of Ref.\
\cite{kr1} in terms of our framework  by using the coherent states
(\ref{coherent}) and quantize the area in its $\lambda$-gauge
fixed form:

 The angular component of the metric tensor is given by
$|E^1|\md\Omega^2$, which leads to the classical expression
$A(x)=4\pi |E^1(x)|$ for the area of a $S^2$-orbit intersecting
the radial manifold $B$ in the point $x$ (in a spherically
symmetric theory these are the only surfaces whose area can be
defined). Writing $$A(x)=4\pi |E^1(x)n_i\,n^i| =4\pi
|E_i(x)\,n^i|$$ we can readily quantize it on a gauge fixed edge
(projected from a $SU(2)$-edge of spin $j$) containing $x$ by
using
$$ n^i\hat{E}_i(x):=
\frac{\hbar\kappa\iota}{4\pi}\ket{m,\vec{n}}_j
n^i\,J_i\bra{m,\vec{n}}_j= \frac{\iota
l_P^2}{4\pi}m\ket{m,\vec{n}}_j\bra{m,\vec{n}}_j $$
where $J_i$ is the angular momentum operator acting on the
coherent state.

However, this quantization depends on what quantum number $m$ we
choose for the projection by means of the coherent state. We can
justify the choice $m=\pm j$ by demanding that we should be able
to recover the spin of the edge uniquely from the projected data.
The simplest way of doing so is given by such a selection of
$m(j)$, namely  choosing $m=j$ is analogous to the extremization
used in Ref.\ \cite{kr1}. In this way we get the spectrum
$$\frac{1}{2}\iota\, l_P^2\,\N_0$$ for the operator, analogously
to Ref.\ \cite{kr1}.

However, in doing so we have used a partial gauge fixing which, as
discussed above, is inappropriate in the non-degenerate sector
because of the Higgs vertices.  If $x$ is a Higgs vertex then we
cannot use this operator because we cannot project at that point
to a $U(1)$-spin network.  This quantization is appropriate only
in the degenerate sectors discussed in Ref.\ \cite{bo1}.

We  now (finally!) quantize the area in the non-degenerate sector
by using $SU(2)$-gauge invariant spin networks and at the same
time undoing any $\lambda$-gauge fixing.  We begin by rewriting
the area into the form \be A(x)=4\pi |E^1(x)|=4\pi\sqrt{(E^1)^2\,
n^i\,n_i}=4\pi\sqrt{E^iE_i}\ee which is  similar to the area
$$A(S_x)=\int_{S_x}\md^2\sigma\sqrt{E^iE_i}$$ of the orbit $S_x$
intersecting $B$ in $x$ in the non-symmetric theory.

 From now on
we can proceed analogously to the quantization of the area
operator in the non-symmetric theory \cite{asle}. As discussed in
the last section, $E_i$ gets quantized to $$\frac{\iota
l_P^2}{4\pi
  i}\fd{A^i}$$ which acts on the $SU(2)$-holonomy $h_e={\cal
  P}\exp\int_e\md xA^i\tau_i$ along the edge $e\colon [0,1]\to B$.

  In order  to
consider a general point $x$ which can be a Higgs vertex we assume
that each edge containing $x$ starts in $x$. We  then have two
outgoing radial edges, one oriented like $B$ itself which we
denote as $e_+$ and one oriented oppositely to $B$ which we denote
as $e_-$, and possibly a Higgs vertex in $x$ which does not depend
on $A^i$ and which can be understood as representing edges
tangential to the surface $S_x$.  By applying the functional
derivative $\delta/\delta A^i(x)$ to a cylindrical function
$f_{\gamma}$ with $\gamma$ containing the edges $e_+$, $e_-$ and
the Higgs vertex $x$ we get
\begin{eqnarray}
 \hat{E}_i(x)f_{\gamma} & = & \frac{\iota l_P^2}{4\pi i}
  \sum_{\epsilon\in\{+,-\}}\int_{e_{\epsilon}}
  \md y\delta(x,y)\tr\left((\tau_i\,h_{e_{\epsilon}})^T
  \pd{h_{e_\epsilon}}\right)f_{\gamma}\\
 & = & \frac{\iota l_P^2}{4\pi}\sum_{\epsilon\in\{+,-\}}
  \frac{\epsilon}{2} J_{e_{\epsilon}}^if_{\gamma}~~.\nonumber
\end{eqnarray}
Here $J_e^i=-iX_e^i$ is given by the $i$-th component of the right
invariant vector field on $SU(2)$.  This leads to the area operator
\begin{equation}
 \hat{A}(x)=\frac{1}{2}\iota\, l_P^2\sqrt{(J_{e_+}-J_{e_-})^2}=
  \frac{1}{2}\iota
  l_P^2\sqrt{2J_{e_+}^2+2J_{e_-}^2-(J_{e_+}+J_{e_-})^2}~,
\end{equation}
with eigenvalues
\begin{equation}\label{spectrum}
 \frac{1}{2}\iota \,l_P^2\sqrt{2j_+(j_++1)+2j_-(j_-+1)-j_v(j_v+1)}~.
\end{equation}
Here the edges $e_+$ and $e_-$ carry the spin $j_+$ and $j_-$,
respectively, and $j_v$ labels the vertex contractor: If $x$ is a
Higgs vertex the associated Higgs point holonomy labeled by a spin
$j$ can be visualized as a loop with spin $j$ based in $x$. This
is in accordance with the Gau{\ss} constraint which can be
regularized to a sum of invariant vector fields containing a left
($J_H^{(L)}$) and a right ($J_H^{(R)}$) invariant one for the
Higgs field \cite{bo1}: $$
J_{e_+}+J_{e_-}=:J_{e_v}=J_H^{(L)}-J_H^{(R)}. $$ Thus $x$ becomes
a 4-vertex whose contractor can be determined by splitting the
vertex into two 3-vertices with a new edge $e_v$ connecting the
edges $e_+$ and $e_-$ with the Higgs loop. It is labeled by the
spin $j_v$ appearing in the eigenvalue of the area operator. Of
course, the Higgs loop as well as the $j_v$-edge have no spatial
extension in the manifold $B$. Because the Higgs field contributes
by left and right invariant vector fields leading to the loop the
spin $j_v$ can only be integer valued. This fact has an immediate
consequence on the topology-dependence of the area operator
discussed in the next section. Here we note that in an appropriate
topology of $\Sigma$ the area spectrum is given by all values of
the form (\ref{spectrum}) where $j_v\in\N_0$ and
$j_+\in\frac{1}{2}\N_0$ are arbitrary whereas $j_-$ is restricted
by $|j_+-j_v|\leq j_-\leq j_++j_v$.  In general, however, the
topology can impose restrictions on the possible values of the
form (\ref{spectrum}) occurring in the area spectrum.

\section{Discussion}

The above area operator in the spherically symmetric sector which
we obtained by restoring the full $SU(2)$-gauge invariance
resembles the one in the non-symmetric theory. The only, however
crucial, difference comes from the simpler topology of
one-dimensional graphs. Therefore we have no sum over vertices
lying on the surface, but only one vertex which represents the
whole surface. This difference influences the area spectrum
considerably:  Disregarding vertex contributions we get the
spectrum \be A(j)= \iota
\,l_P^2\sqrt{j(j+1)}~,~~j\in\frac{1}{2}\N_0~,\ee which is only a
small subset of the corresponding spectrum (2) in the
non-symmetric theory. In particular, for large $j$ the spectrum
becomes not dense, but equidistant, and in the large $j$ limit we
obtain the spectrum (1) of the horizon area described in the
introduction.

Thus we have shown that loop quantum gravity in its spherically
symmetric sector reproduces for large spins, i.e.\ in the assumed
semiclassical regime, the older results while it leads to
corrections for small $j$, i.e.\ at the Planck scale.

As in the case of the area operator in the non-symmetric theory
the spectrum of the spherically symmetric one depends on the
topology of space. Here any surface whose area we can measure in a
spherically symmetric theory has, of course, the topology of
$S^2$. However, there are essentially two possible space
topologies: A topology with two (or more) boundary components and
second homology $H_2(\Sigma)=\Z$, and one with a single boundary
component and $H_2(\Sigma)=0$ (we regard spatial infinity as a
boundary). In the first case we have two physical realizations:
The wormhole topology $B=\R$, $\Sigma=\R\times S^2$ represents a
spacelike hypermanifold in the Kruskal extension of Schwarzschild
(or Reissner-Nordstr{\o}m) space-time and has two boundary
components at $\pm\infty$, whereas the topology $B=\R_+$,
$\Sigma=\R^3\backslash\{0\}$ can be seen as simulating an
external, non-dynamical gravitational source sitting in the
origin, i.e.\ in one of the two boundary components of $\Sigma$.

The topology with only one boundary component is given by
$B=\R_+\cup\{0\}$, $\Sigma=\R^3$ and has only the boundary at
spacelike infinity. Here we have to treat the symmetry center in
$0$ along the general framework of symmetry reduction. The
isotropy subgroup is $F=S=SU(2)$ and the homomorphism
$\lambda\colon F\to G$ is either $\overline{\lambda}_0\colon
g\mapsto 1$, $\overline{\lambda}_1^{(\prime)}\colon g\mapsto
g\cdot \{\pm 1\}\in G\backslash\{\pm 1\}\cong SO(3)$ or
$\overline{\lambda}_1\colon g\mapsto g$ (up to conjugacy) for all
$g\in S$. This can be seen from the fact that the kernel of
$\lambda$ is an invariant subgroup of $S$ which can only be $S$,
$\{\pm 1\}$ or $\{1\}$. By continuity, in the rest of $B$ we must
use the homomorphism $\lambda_k$ if in $0$ we use
$\overline{\lambda}_k$ ($\overline{\lambda}_1^{(\prime)}$ and
$\overline{\lambda}_1$ make no difference). This shows that we
have only the possibilities $k=0,1$ if the symmetry center is
contained in $\Sigma$. The two possibilities lead to manifestly
invariant connections ($k=0$) and to connections invariant up to
gauge ($k=1$).  In all these cases there can be no Higgs field in
$0$, which is in accordance with the fact that the Higgs field
represents components of an invariant connection tangential to the
$S$-orbits which is a single point in the case of $0$. An
immediate consequence of this fact is that $F=S$, which implies
${\cal L}F_{\perp}=\{0\}$ (in the Cartan-Killing metric) and
therefore the Higgs field which is a map $\phi\colon{\cal
L}F_{\perp}\to{\cal L}G$ vanishes.

We can now consider the implications of these considerations as to
gauge invariant spin networks. The crucial observation is that in
a Higgs vertex the spins of the neighboring edge holonomies can
differ only by an integer value because the spin $j_v$ mentioned
above is integer-valued due to the Higgs loop in the vertex. If
$\Sigma$ has two boundary components  we do not have to enforce
$SU(2)$-gauge invariance in the two boundary points of $B$ and the
edges can be either all integer valued or all half-integer valued.
This leads to the full spectrum (\ref{spectrum}) given above.
However, if $0\in B$ corresponds to the symmetry center, i.e. an
inner point of $\Sigma$ implying that there can lie no Higgs
vertex, we have to impose $SU(2)$-gauge invariance in $0$. The
edge incident in $0$ can only have spin $0$ which implies that the
other edge spins can only be integer valued. This fact allows only
a subset of (\ref{spectrum}) as area spectrum.

As a last remark we note that the spectroscopy for spherically
symmetric black holes (cf.\ e.g.\ \cite{be2}) is unaltered by our
area spectrum (13) because it becomes uniformly spaced for large
$j$, like the spectrum (1). This is not the case for the full area
spectrum (2) of the non-symmetric theory which becomes almost
continuous.

The large discrepancy between these two spectra may be understood
as a line splitting due to a broken symmetry. Because of the
discrete structure of space made explicit by a spin network
spherical symmetry is strongly broken by a state in the
non-symmetric theory. As is well known in quantum theory, breaking
a symmetry can lead to a splitting of levels which are degenerate
before breaking the symmetry. In our case the degeneracy of the
levels of a black hole is expected to be very huge, growing
exponentially with $j$ (see section 3 of Ref.\ \cite{ka5} and the
literature mentioned there). Splitting of these strongly
degenerate levels by a broken symmetry may lead to an almost
continuous spectrum as observed in the non-symmetric theory. This
observation may also open up a new way to calculate the degeneracy
of the energy levels of black holes in loop quantum gravity.

\section*{Acknowledgement}

M.B.\ thanks the DFG-Graduierten-Kolleg ``Starke und
elektroschwache Wechselwirkungen bei hohen Energien" for a PhD
fellowship.

\end{document}